\documentclass[aps,preprint,showpacs,floatfix,pra]{revtex4-1}
\usepackage{bbding}
\usepackage{epsfig}
\usepackage{graphicx}
\usepackage{dcolumn}
\usepackage{bm}

\usepackage{amssymb}
\usepackage{amsmath}       
\begin{document}
\title{ Accurate control of a Bose-Einstein condensate by managing the atomic interaction}

\author{L. Morales-Molina}
\affiliation{Departamento de F\'isica, Facultad de F\'isica, Pontificia Universidad Cat\'olica de Chile, Casilla 306, Santiago 22, Chile}
\author{E. Ar\'evalo}
\affiliation{Max-Planck-Institut f\"ur Physik Komplexer
Systeme, N\"othnitzer Str. 38, 01187 Dresden, Germany}

\begin{abstract}
We exploit the variation of the atomic interaction in order to move ultra-cold atoms across an AC-driven periodic lattice. By breaking relevant symmetries, a gathering of atoms is achieved. Accurate control of the gathered atoms' positions can be demonstrated via the control of the atomic localization process.
The localization process is analyzed with the help of the {\it nonlinear Floquet states} where the Landau-Zener tunneling between states is observed and controlled.
Transport effects in the presence of disorder are discussed.
\end{abstract}

\date{\today}
 \maketitle

\section{Introduction}

 In nonlinear discrete lattices, spatial discreteness and nonlinearity constitute the two main ingredients necessary for the existence of localized excitations.
The existence and dynamics of these excitations can be understood from the analysis of their energy spectrum. In this spectrum, the localized states lie in gaps between bands of extended states.
 
The spectrum, and therefore the localized states, depend very much on the strength of the nonlinearity since the position and width of the bands and gaps vary with this parameter. 
In this respect, the nonlinearity strength can be used as a control parameter for tuning localized excitations.
An ideal experimental testing ground for the application of this type of control is that of a Bose-Einstein condensate (BEC) trapped in a deep optical lattice, where much of the nonlinear and discrete effects can be probed in optical lattices \cite{Opt-Latt}.

Further control of matter waves (BEC) in optical lattices has been demonstrated with the use of fields. AC-fields can modify the properties of the matter waves, by adjusting the parameters of the fields \cite{Matterwaves}. This is an intense and growing area of research, where examples of striking phenomena, such as  transitions of a superfluid to a Mott-Insulator \cite{Mott-transit} and generation of directed transport \cite{PRAresonan,eplresonan}, have found an experimental footing \cite{Lignier,ScienceTobias}. 

In this respect, a combination of AC fields together with management of nonlinearity opens new avenues for exploration of the new phenomena in BECs trapped in optical lattices.

In this work, we consider a general management of the atomic interaction for a BEC in presence of an AC driving field. Manipulation of the atomic interaction is usually referred to ``management of Feschbach resonances'' \cite{kevre}. At these resonances, the scattering length undergoes large fluctuations with values going from positive to negative as the magnetic fields are tuned \cite{ketterle}, thus opening many possibilities for the control of the atomic interaction. 
``Management'' here refers to manipulation of DC and AC parts of the atomic interaction. The AC part is locally applied to induce motion of the atoms towards the perturbed sites. In the light of new experiments, local variation of the atomic interaction is feasible by spatially modulating the interaction strength on a short-length scale \cite{NaturePhys}. 
Here, rather than focusing on the localization process itself, we analyze how to get transport of particles out of the localization process; i.e., how to transport cold atoms within optical lattices by controlling the location of the localization sites.

\section{Model}

Consider a lattice with periodic boundary conditions under the action of an AC force. The dynamics can be described by the equation
 
\begin{equation}\label{Eq:lattice}
i \frac{d \psi_{n}}{dt}=  E_{n}\psi_{n}+C (\psi_{n+1}+\psi_{n-1})+g\,\psi_{n} N _{n} -\psi_{n}F_{n}(t), 
\end{equation}
 where $E_{n}$ is the energy at the site $n$. $N_{n}= N |\psi_{n}|^2$ is the population of atoms at the $i^{th}$ site. $N$ is the total number of atoms, that for convenience is normalized to $1$. For the sake of dimensionless units, we set $\hbar=1$.
 $g$ accounts for the interaction strength between atoms, which is a function of the s-wave scattering length. This in turn can be tuned by means of
magnetic fields \cite{ketterle}. More recently, control of the scattering length has been realized with a laser \cite{NaturePhys}, allowing more accuracy and a less significant loss of atoms. 
$F_{n}(t)$ in Eq.\ref{Eq:lattice} is a periodic function that mimics a flashing potential \cite{ScienceTobias} for a discrete lattice. This can be written as
\begin{equation}\label{Eq:field}
F_{n}(t)=(-1)^{n}f(t)\equiv (-1)^{n}h\sin(\omega t),
\end{equation}
  where $\omega$ is the driving frequency  and  $h$ is the amplitude.  The term $(-1)^n$, accounts for the spatial periodicity of the potential. Eq.(\ref{Eq:lattice}) with (\ref{Eq:field}) satisfies periodic boundary conditions.

In Eq.(\ref{Eq:lattice}) the  atomic interaction strength $g$ can include CD and AC terms whose manipulation  is usually referred to as ``management of Feschbach resonances (MFR)'' \cite{kevre}.  
 MFR  has proven to be an effective tool, on the mean field level, for manipulation of cold atoms  \cite{kevre}. It has also been suggested as way to control the tunneling process of a discrete number of atoms \cite{GongCDT}.

Notice, moreover, that the control of atomic population imbalances in a double well potential has been realized by varying parameters in the system including the atomic interaction strength between atoms \cite{rapid}. A similar kind of control has been proposed for controlling the wavepacket spreading in a disordered lattice \cite{Vicencio}. Upon varying the ramping speed of a linear time-dependent strength of the nonlinear term, the authors find optimum localization in the lattice in presence of disorder.

From the perspective of a cold-atom system, localization, (namely gathering of atoms at few sites) implies a motion of atoms across the lattice, meaning a transport of particles. Notice that the process here is different from that of moving solitons, where the motion is achieved by kicking localized matter waves \cite{Edward}.
 Thus, controlling the location of the localization sites allows us to control the transport of the atoms.
With this aim, we consider DC and AC terms for $g$, namely
\begin{equation}\label{Eq:interaction}
 g=g_{0}+U \sin(\omega t+\theta),
\end{equation}
 where $\theta$ is a phase.
 It is expected that the periodic field in Eq.(\ref{Eq:field}), along with management of the atomic interaction, may help  to control localization of atoms at specific sites of the lattice.

\begin{figure}
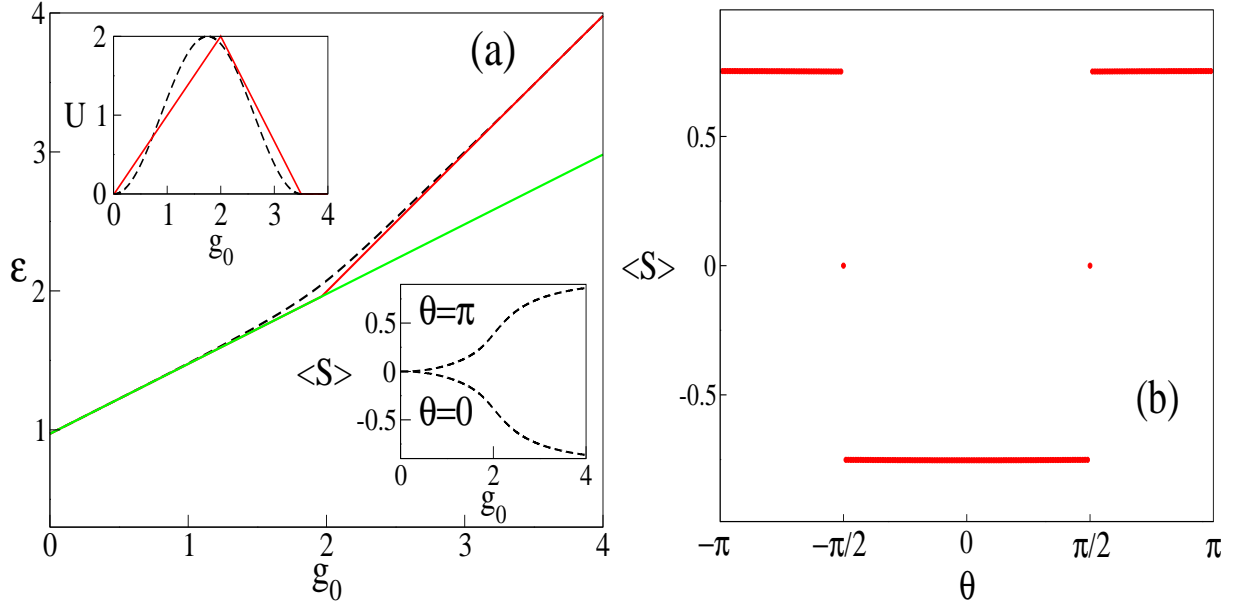

 \begin{center}
\begin{tabular}{lc}
\includegraphics[width=8cm,height=8cm]{fig1a.eps}
\includegraphics[width=8cm,height=7.9cm]{fig1b.eps}
\end{tabular}
\caption{(Color online) (a) Quasienergy $\varepsilon$ vs. $g_{0}$ for the two modes system [Eqs. (\ref{Eq:dimer1}) and (\ref{Eq:dimer2})]. Green and red solid curves: $U=0$; black dashed curve: $U^{max}=2$. 
Upper inset: Examples of pulse shape: $U$ vs. $g_{0}$: two-slope (solid line) and Sine-square (dashed line). Lower inset:  Average of population imbalance $\langle S \rangle$ vs. $g_{0}$ for $U^{max}=2$. (b) $\langle S \rangle $ vs. phase $\theta$, $U^{max}=2$. The average $\langle ... \rangle$ is realized over a period $T=2\pi/\omega$. Computations are performed with the sine-square shape depicted in the upper inset of (a).
The parameters are $C=1$, $\omega=2 \pi$, $h=1$, $E=0$, $g_{0}= 3$.}
\label{Fig:Spectrum-nonlinear}
\end{center}
\end{figure}

\subsection{Two-mode system}

 To gain an insight into the dynamics, it is convenient first to analyze the dimer lattice, which can be written as 

\begin{eqnarray}\label{Eq:dimer1}
i \frac{d \psi_{1}}{dt}&= & E\psi_{1}+C \psi_{2}+g\,\psi_{1} N _{1} +\psi_{1} f(t),
\\
i \frac{d \psi_{2}}{dt}&=&-E\psi_{2}+C \psi_{1}+g\,\psi_{2} N_{2}-\psi_{2} f(t).
\label{Eq:dimer2}
\end{eqnarray}

Here the energies are set as $E_{1}=E$ and $E_{2}=-E$, where $E$ can be seen as a bias.

 Eqs. (\ref{Eq:dimer1}) and (\ref{Eq:dimer2}) with $E=f(t)=0$  have been used to describe the self-trapping transition of two weakly coupled BEC \cite{smerzi}, 
  This phenomenon appears when the nonlinearity 
  exceeds a critical value $g_{c}$ \cite{kladko,smerzi}.

To understand this phenomenon in the presence of an AC field $f(t)$,  it is convenient to study the eigenfunctions of the extended Hilbert space of $T$-periodic functions, similar to that implemented in the linear regime (see Ref. \cite{Sambe}). 
Conventional Floquet theory fails in the presence of a nonlinear dependence on the wave functions. Nonetheless, Floquet states of the linear system 
can be traced into the nonlinear domain and new periodic orbits are found, which are then called Nonlinear Floquet states \cite{holthaus0,holthaus,njp,rapid}. The periodic solutions in the linear regime $g=0$ fulfill the relation
$ |\psi_{\beta}(t)\rangle= e^{-i \textstyle \varepsilon_{\beta} t} |\phi_{\beta}(t)\rangle$, where $|\phi_{\beta}(t+T)\rangle=|\phi_{\beta}(t)\rangle$ and $\varepsilon_{\beta}$ is the quasienergy value.

To characterize the localization of atoms in a two-mode system, we use the quantity $S=|\psi_{1}|^2-|\psi_{2}|^2$, which is the atom population imbalance  between the two sites. 

Equations (\ref{Eq:dimer1}) and (\ref{Eq:dimer2}) have two Floquet states for $g=0$ with zero population imbalances, namely $S=0$.
 When increasing $g$ above some $g_{c}$, one of the two states bifurcates into three new states, as shown in Fig.\ref{Fig:Spectrum-nonlinear}a.
Two of these new states are degenerate, with opposite population imbalances. These are impossible to reach by continuation from the linear regime, because of the breaking of the adiabatic condition at the branching point \cite{rapid}.
Let us now analyze the symmetries of Eqs. (\ref{Eq:dimer1}) and (\ref{Eq:dimer2}) in the presence of MFR (see Eq.(\ref{Eq:interaction})).

If $U=0$ in Eq.(\ref{Eq:interaction}), Eqs. (\ref{Eq:dimer1}) and (\ref{Eq:dimer2})  become invariant under the transformations:
\begin{equation}\label{Eq:Symme1}
S_{1}: \psi_{1}\rightarrow \psi_{2}, \,\ t \rightarrow t+T/2, 
\end{equation}
and 
\begin{equation}\label{Eq:Symme2}
S_{2}: \psi_{1}\rightarrow \psi_{2}, \,\ t\rightarrow -t, \mbox{ complex conjugation,}
\end{equation}
where $S_{1}$ corresponds to a generalized permutation symmetry.

On the contrary, for $U\ne0 $, $S_{1}$ is always broken, and  $S_{2}$ is broken 
if $\theta \ne \pm \pi/2$.

Thus, taking Eq.(\ref{Eq:interaction}) in Eqs. (\ref{Eq:dimer1}) and (\ref{Eq:dimer2}) with $\theta\ne \pi/2$, breaks both $S_{1}$ and 
$S_{2}$ symmetries.

To follow states from the linear regime with zero population imbalances, it is convenient to modulate the AC nonlinear term $U$ in Eq. (\ref{Eq:interaction}) as a function of $g_{0}$, namely $U(g_{0})$. In the following, we assume   
$U(g_{0})$ has a pulse shape.
As a result of this modulation, a new energy branch is created off the branching point that continuously join states with zero or little population imbalance to those states of strong population imbalances in Fig.\ref{Fig:Spectrum-nonlinear}a.
The pulse shape appears not to be very important, since the plots of the quasienergy curves obtained for pulses with a two slope- and sine-square shapes (see upper inset of Fig.\ref{Fig:Spectrum-nonlinear}a)  show no significant difference. From here onwards, we 
take on the two slope-shape. 
To reach those states with strong population imbalances, we simply ramp  $g_{0}$ linearly  in time, viz.    

\begin{equation}
 g_{0}(t)\equiv  \alpha t,
\end{equation}
where $\alpha$ is the speed of variation or ramping rate.

\begin{figure}
 \begin{center}
\includegraphics[width=7.5cm,height=6.5cm]{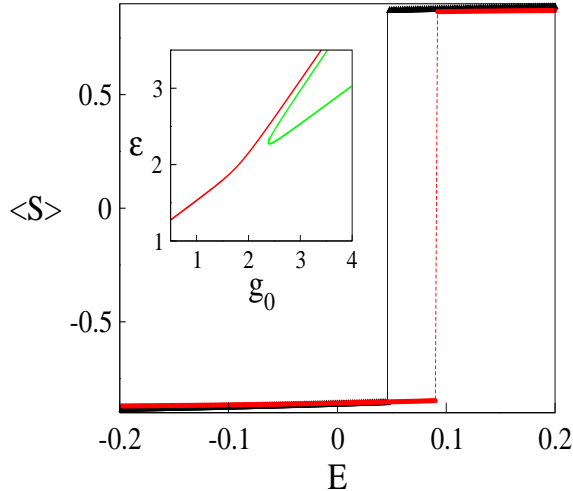}
\caption{(Color online) Average population imbalance $\langle S \rangle$ vs. bias E for $U^{max}=2$, $h=1$ (black), $h=2$ (red); $g_{0}=4$, $\theta=0$. We compute in the limit $\alpha=0$. Computations are performed with the sine-square shape depicted in the upper inset of Fig.\ref{Fig:Spectrum-nonlinear}(a). Inset: Quasienergy $\varepsilon$ vs. $g_{0}$. The parameters are $C=1$, $h=1$, $E=0.1$, $U=0$.}
\label{Fig:Popul-vsBias}
\end{center}
\end{figure}

Interestingly, changes of the phase $\theta$ creates states with opposite population imbalances, as shown in lower inset of Fig.\ref{Fig:Spectrum-nonlinear}a. A full scan of the phase shows that the population imbalance becomes zero for $\theta=\pm \pi/2$ only (see Fig. \ref{Fig:Spectrum-nonlinear}b). This is in agreement with the symmetry analysis from Eqs. (\ref{Eq:Symme1}) and (\ref{Eq:Symme2}).
States with opposite population imbalances can therefore be selectively targeted when slowly changing $g_{0}$. 
Moreover, the smoothness of the new energy branch in the vicinity to the bifurcation point allows a significant increase in the ramping speed $\alpha$. This is because a sharp variation in quasienergy is usually interpreted as a strong interaction with a very near eigenstate 
\cite{eplZener}. If the energy variation is smooth, high $\alpha$ values can be taken into account. However, for too high $\alpha$, the system may hop from one state to other states with zero or little population imbalance. This implies   a loss of localization. This jump brings to mind the Landau-Zener tunneling observed in the linear regime \cite{eplZener}. 

Furthermore, Eqs. (\ref{Eq:dimer1}) and (\ref{Eq:dimer2}) can describe the dynamics of opposite momenta for a condensate \cite{njp}. A generalization to an extended lattice, together with Landau-Zener tunneling between states of different momenta, may significantly change the transport of the atoms \cite{njp}.

 The above technique comprises a rather general MFR (Eq.\ref{Eq:interaction}) for the atomic interaction.     
Thus, we can partly summarize the main effects on the spectrum of quasienergies  when the above generalized management of the atomic interaction is applied:
 first, it creates a new quasienergy path that circumvents the branching point, 
second, the new path allows a rapid passage, and third, positive or negative strong population imbalances is reached upon choosing the phase $\theta$.

So far, the analysis has been focused for a zero bias, i.e. $E=0$ in Eqs.(\ref{Eq:dimer1}) and (\ref{Eq:dimer2}). For a nonzero bias $E\ne 0$, degeneracies are lifted  as shown in inset of Fig.\ref{Fig:Popul-vsBias}. In this case, localization can also be achieved by slowly changing $g_{0}$ in time.
  In the same spirit as the functioning of a ratchet system \cite{HanggiRatchet}, here we are interested in how localization is induced by means of AC fields and, more importantly, how this works against a bias.    
Our results show that the induced localization  appears to work against a bias as shown in Fig.\ref{Fig:Popul-vsBias}. An averaged nonzero population imbalance is induced  for $E=0$, whose effect survives for finite values of $E$. The offset is shifted to the right as the amplitude of the field $f(t)$, $h$, increases. Likewise, one can induce the same effect to the left by changing $\theta$ in $f(t)$, Eqs.(\ref{Eq:dimer1}) and (\ref{Eq:dimer2}).

\subsection{Lattice}

 \begin{figure}
 \begin{center}
\includegraphics[width=8.5cm,height=7.5cm]{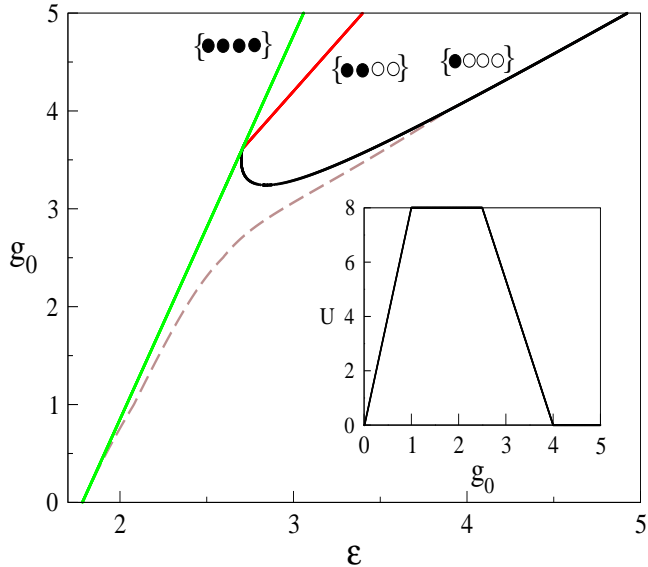}
\caption{(Color online) Nonlinearity strength $g_{0}$ vs. quasienergy $\varepsilon$ for $n=4$. $h=4$. Dashed line is a quasienergy state that results from continuing the linear Floquet state into the nonlinear domain, in the presence of an AC pulse of the nonlinearity strength, with the profile shown in the inset. Inset: Pulse profile  $U$ vs. $g_{0}$, $\omega=4\pi$, $C=1$.}
\label{Fig:spec-4modes}
\end{center}
\end{figure}
 
 As mentioned above, states bifurcate in new branches, representing new solutions, as the nonlinearity strength increases.
Hence, in order to describe the many possible solutions that appear in the system of Eqs. (\ref{Eq:lattice}) and (\ref{Eq:field}), as the nonlinearity increases, a shorthand description for each solution of the branch is convenient. Such analysis is usually realized in the limit of $g \rightarrow \infty$, where one can  estimate  the asymptotic solution \boldmath $\psi $ \unboldmath $\equiv (\psi_{1},\psi_{2},...)$ for every branch \cite{eilbeck}.

Rather than determine the solution \boldmath $\psi $ \unboldmath $\equiv (\psi_{1},\psi_{2},...)$, here we are interested in the  modulus-squared of the wave function in the sites, i.e,  $|\psi_{i}|^2$. That is, the number of atoms localized at each site $i$. Thus, we take a modified shorthand representation, where now ``$\bullet$'' accounts for the occupied sites with a $1/K$ fraction  and ``$\circ$'' for a zero fraction of atoms of the corresponding state in the limit $g \rightarrow \infty$. Here $K$ is the number of nonzero $\psi_{i}$. 

In the following, we consider a four-mode lattice. In this case, Eq.\ref{Eq:lattice} for four sites remains  invariant under permutations of the indices $1\rightarrow 3$ and $2\rightarrow4$. Likewise, symmetries given by Eqs. (\ref{Eq:Symme1}) and (\ref{Eq:Symme2})  are also applicable. 
Figure \ref{Fig:spec-4modes} shows a section of the quasienergy spectrum vs. nonlinearity strength. Of 
particular interest are the states   $\{\bullet\, \bullet\, \bullet\, \bullet\}$  with equal distribution of atoms across the lattice and  $ \{\bullet\,\circ\,\circ\, \circ \}$, with all the atoms localized in one site.
\begin{figure}
 \begin{center}
\includegraphics[width=12cm,height=6.5cm]{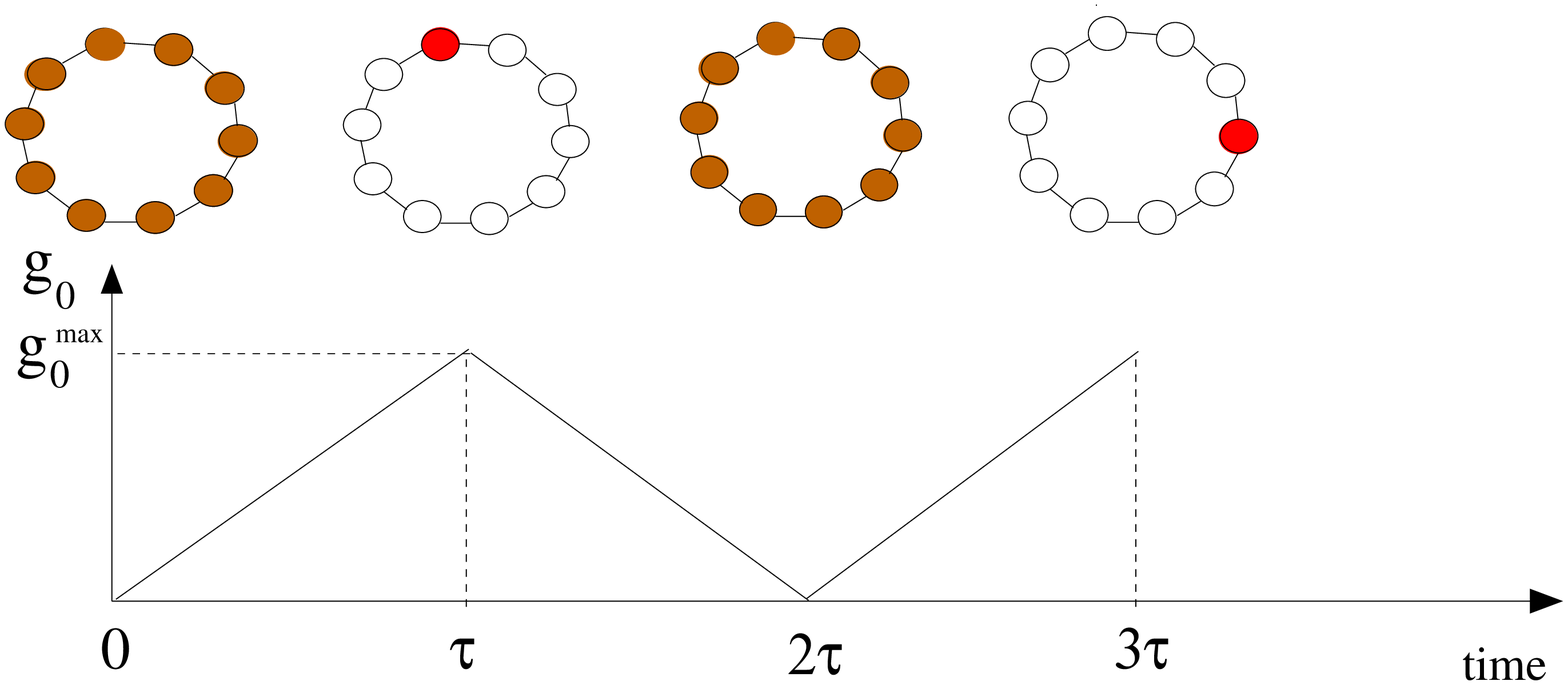}
\caption{(Color online) Lower: Sketch of the ramping process. Upper: Distribution of atoms in a ring of connected sites  at the beginning and end of every ramping process. (For meaning of colors read text). }
\label{Fig:sketch}
\end{center}
\end{figure} 
 
We consider now the MFR [Eq.(\ref{Eq:interaction})] which creates as mentioned above, a new branch that join the zero-population-state with that of very high population imbalance as  $g_{0}$ increases (Fig.\ref{Fig:spec-4modes}, dashed line).

Interestingly, as the pulse  of the AC nonlinearity strength $U(g_{0})$ is applied on two sites only, the number of atoms tend to localize in one of these two sites when ramping $g_{0}$. Meaning that a motion of atoms from all the other  sites to this new populated site has taken place. This can be fully controlled with the phase $\theta$ [see Eq.(\ref{Eq:interaction})]. 
In this respect, the combined action of  ramping $g_{0}$ along with an AC perturbation of $g $ [see Eq.(\ref{Eq:interaction})] can, in principle, help to control the movement of atoms along the lattice. Figure \ref{Fig:sketch}  schematically shows a method to move atoms from one lattice site to another. The lower graphic depicts two stages in time, each one consisting of two ramping process of $g_{0}$, which are described as follows: As time proceeds $g_{0}$ increases until it reaches a maximum value $g_{0}^{max}$ at $\tau=g_{0}^{max}/\alpha$, where a localization of atoms takes place.  Afterwards $g_{0}$ is ramped down in time until $g_{0}=0$ at $2\tau$ (at this point the number of atoms are equally distributed). At the second stage, the AC perturbation is applied on two sites, different from the prior ones, thus making  atoms gather in a different new lattice location. In doing so, we have effectively achieved moving atoms from one lattice site to another at time $3\tau$.
The entire process of  moving atoms to different sites in a circular ring is depicted in the upper part of Fig.\ref{Fig:sketch} by a sequence of drawings. Each drawing represents a configuration of localization of atoms in the ring at the end of every ramping process. The red color on one site stands for a large concentration of atoms at this site, whereas absence of color represents a zero or very low number of atoms. Drawings with brown color at every site means an equal distribution of atoms. Notice that this process can be repeated over and over, to shift atoms to any target site of the lattice.

A remaining open issue is why the atoms initially distributed in the whole lattice gather at a single site. Though it is clear that motion takes place via a tunneling process of atoms across the lattice, the way the atoms tunnel may involve not only single-order tunneling (tunneling of a single particle) but also higher orders of tunneling (tunneling of bound particles) \cite{ScienceTunneling}.

\begin{figure}
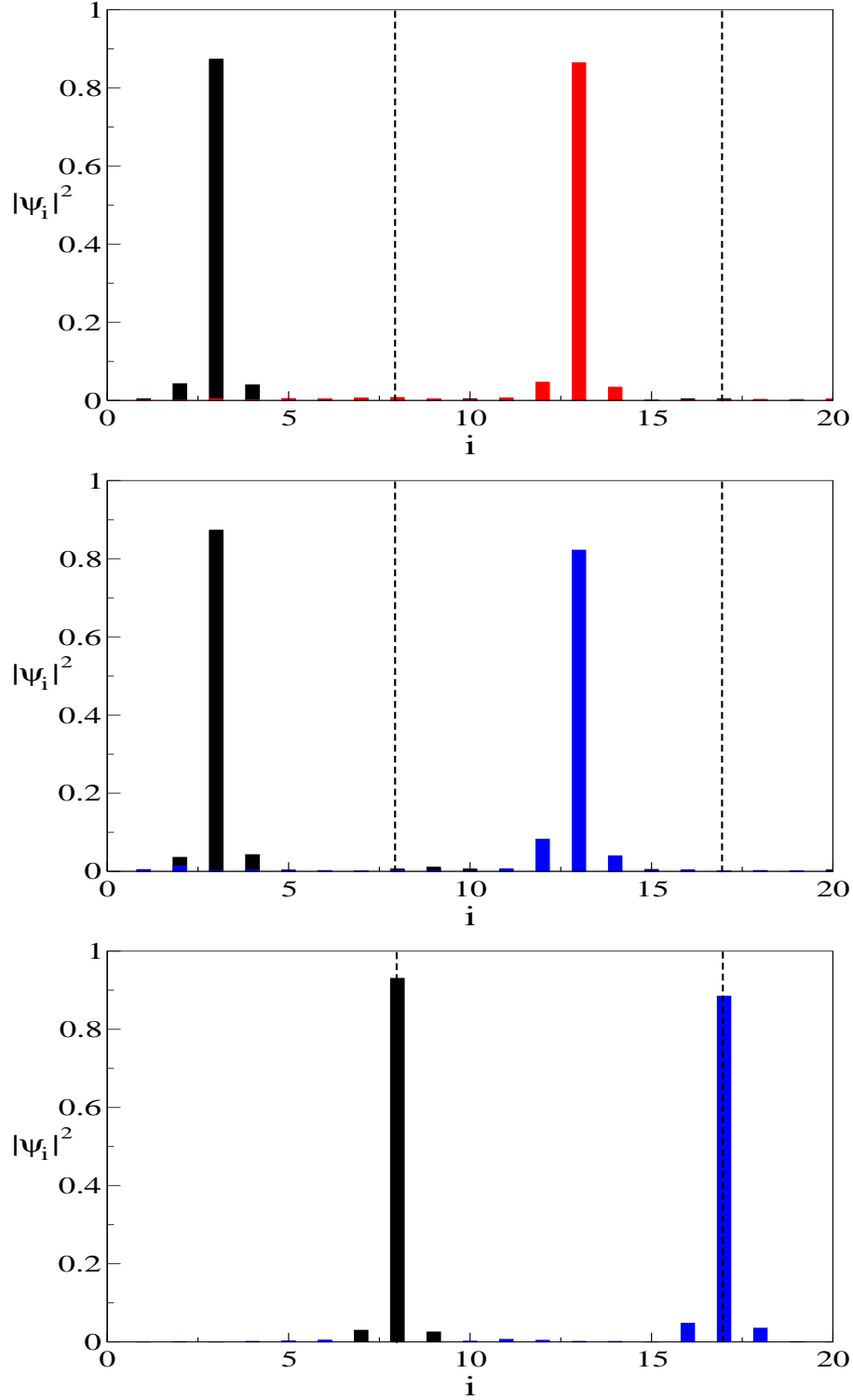

 \begin{center}
\begin{tabular}{lc}
\includegraphics[width=12cm,height=6.5cm]{fig5a.eps}\\
\includegraphics[width=12cm,height=6.5cm]{fig5b.eps}\\
\includegraphics[width=12cm,height=6.5cm]{fig5c.eps}
\end{tabular}
\caption{(Color online) Atoms density $|\psi_i|^2$ vs. location site $i$. 
(Upper) $Q=0$; (Middle) $Q=0.025$; (Lower) $Q=0.05$; The dashed lines indicate the impurities location at the sites $i=8,17$. (see text for details on the localization sites) $C=1$, $\omega=4\pi$, $h=4$, $\alpha_{f}=0.1$, $\theta=\pi$, $\alpha=5\times 10^{-3}$, $g_{0}^{max}=5$. Computations are performed with the $U(g_{0})$ shape depicted in the upper inset of Fig.\ref{Fig:spec-4modes}. }
\label{Fig:distribution}
\end{center}
\end{figure}  

Up to now, the analysis has been focused on the manipulation of atoms at the level of Floquet states. Clearly, to steer atoms in the lattice, one needs first to populate those states. Population of the Floquet state where atoms appear equally distributed across the lattice is realized by ``slowly'' turning on the periodic field, using a uniform distribution of atoms as the initial condition. That is, we ramp $f(t)$ amplitude (Eq.\ref{Eq:field}), $h$, from 0 to its maximum value. This process is carried out at a speed  $\alpha_{f}$ that satisfies the constraint $\alpha_{f}/\omega\ll 1$. 

In the following, we consider a lattice of 20 sites. In Fig.\ref{Fig:distribution}  the  atom number distribution is recorded after ramping  up $g_{0}$ during the two consecutive stages, as depicted in  Fig.\ref{Fig:sketch}. Notice the apparent localization of atoms on the sites that are perturbed by the AC pulse $U(g_{0})$ [see Eq.\ref{Eq:interaction}]. In the first stage, the AC pulse is applied on the sites 2 and 3 and localization takes place on the site 3. In the second stage, the AC pulse acting on sites 13 and 14 causes localization  on site 13.
The localization on the perturbed sites depends on $\theta$ similarly as in Fig.\ref{Fig:Spectrum-nonlinear} for the two modes system. 
 
A question that arises is whether this procedure can be applied to much larger lattices, and which limitations exist. Further tests indeed indicate that the movement of atoms is possible for a lattice with a larger number of sites, at the expense of reducing the  ramping speed. 
The larger the lattice becomes, the longer the time is needed for the atoms to move across the lattice and gather together at few sites.

 From the quasienergy analysis, more sites in the lattice implies more Floquet states in the quasienergy spectrum with energies becoming more densely packed and with smaller energy gaps. This increases the probability for Landau-Zener transitions, thus decreasing the control over the atoms.

 We have showed for a two-mode system that localization may take place against a bias.  Lattice impurities cause a similar effect in the energy spectrum to that of a bias for a dimer setting (cf bifurcation in Ref. \cite{Bifurdisorder} with inset in Fig.\ref{Fig:Popul-vsBias}).
So, it is of particular interest to see what  happens with the movement of atoms from one site to any other site in the presence of impurities.

Here, we consider two impurities placed between the initial and final sites of the motion process. The impurities are introduced  by changing the energies at some specific sites, i.e, $E_{n}=Q*(\delta_{n,k}+\delta_{n,m})$, where $\delta_{n,i}$ is the Kronecker's delta function. $k$ and $m$ are the indices of the impurity sites in a lattice that runs from $n=0,...,20$ and $Q$ is the amount of energy at those sites. For low $Q$, movement of atoms takes place as in the homogeneous lattice, whereas for large $Q$, the atoms tend to localize at the impurity sites, shown in Fig.\ref{Fig:distribution}.

We performed similar computations with a random distribution of impurities (not shown). The results observed are similar to those exhibited above, where the atoms tend to gather at impurities with large $E_{n}$.

\subsection{Conclusions}

In summary, we have shown the steering of a Bose-Einstein condensate with a generalized management of the atomic interaction in a nonlinear discrete AC driven lattice.
 Transport of atoms is realized via the control of the atom's localization process. The whole process relies on gathering atoms in different sites of the lattice upon successive ramping move in time of a DC part of the atomic interaction strength.
 
Localization of atoms is achieved upon ramping the atomic interaction assisted by an AC field, together with periodic oscillations of the nonlinear interaction. 
 This process of localization is 
 first explained within a two-mode approximation, where the sign of the population imbalances is controlled by the phase of the periodic functions. This is supported by  a symmetry analysis.
We also show within a two-mode setting that control of the localization, 
 created by the application of AC perturbations, may overcome a bias.

This was later shown to be feasible in the lattice, when movement of atoms was proven to be robust against a low intensity of disorder.

 The control procedure exposed here paves the way for further application in other nonlinear systems, such as nonlinear optics where the nonlinearity strength is modulated in optical waveguides. Likewise, these results can be used for a generic equation like the discrete self-trapping equation with application to fundamental problems, such as dynamics of small molecules \cite{molecules}, dynamics of molecular crystals \cite{Crystaline}, amongst others.
On the other hand, similar management of the atomic motion  may be potentially useful  in the control of momenta states in a condensate and consequently for the transport of ultra-cold atoms in optical lattices. 

\begin{acknowledgments}
The authors are greateful to fruitful comments on symmetries by Sergej Flach and also to suggestions by Joshua D. Bodyfelt.  L.M-M is greateful for the hospitality of the MPIPKS. L.M-M  acknowledges financial support from Mecesup de Chile.
\end{acknowledgments}

\
\end{document}